\documentclass[prb,twocolumn,preprintnumbers,amsmath,amssymb]{revtex4}

\usepackage{graphicx}
\usepackage{dcolumn}
\usepackage{bm}
\usepackage{amsmath}
\usepackage{amssymb}
\usepackage{amsthm}
\usepackage{amsfonts}
\usepackage{enumerate}
\usepackage{latexsym}

\usepackage{hyperref}
\usepackage{tikz}

\begin{document}

\title {Nonperturbative emergence of Dirac fermion in strongly correlated composite fermions of fractional quantum Hall effects }

\author{Yibin Yang$^{1}$}
\thanks{These two authors contribute equally.}
\author{Xi Luo$^{1}$}
\thanks{These two authors contribute equally.}
\author{Yue Yu$^{2,3,4}$}

\affiliation {1. CAS Key Laboratory of Theoretical Physics, Institute of
Theoretical Physics, Chinese Academy of Sciences, P.O. Box 2735,
Beijing 100190, China\\
2.Center for Field
Theory and Particle Physics, Department of Physics, Fudan University, Shanghai 200433,
China \\
3. State Key Laboratory of Surface Physics, Fudan University, Shanghai 200433,
China\\
4. Collaborative Innovation Center of Advanced Microstructures, Nanjing 210093, China}

\begin{abstract}
The classic composite fermion field theory \cite{HLR}  builds up an excellent framework to uniformly study important physical objects and globally explain anomalous experimental phenomena in fractional quantum Hall physics while there are also  inherent weaknesses.  We present a nonperturbative emergent Dirac fermion theory from this strongly correlated composite fermion field theory, which overcomes these serious long-standing shortcomings. The particle-hole symmetry  of Dirac equation resolves this particle-hole symmetry enigma in the  composite fermion field theory. With the help of presented numerical data, we show that  for main Jain's sequences of fractional quantum Hall effects,  this emergent Dirac fermion theory in mean field approximation is most likely stable.
\end{abstract}


\maketitle

\section{Introduction}\label{intro}

 The emergent relativistic fermions are ubiquitous in condensed matter systems \cite{ti1,ti2,gp,volo,weyl}. Most of them are derivatives of the free non-relativistic electrons that are subject to various influences, such as a special periodic potential from the underlying lattice, or flux attaching or spin-orbital couplings, and so on. The derivatives from strongly correlated systems were rarely seen and often immit completely new phenomena or concepts. Majorana fermion in $\nu=\frac{5}2$ fractional quantum Hall effects (FQHE) was an excellent example \cite{GR}. This results in the birth of the concepts of nonabelian fractional statistics \cite{MR} and topological quantum computer \cite{kitaev}.  
 
There was  an ambiguity in introducing Majorana fermion at $\nu=\frac{5}2$. While Majonara fermion is a fully relativistic object \cite{GR}, Moore-Read Pfaffian is the variational ground state wave function of a two-dimensional non-relativistic electron gas in a strong external magnetic field \cite{MR}. The seminal Halperin-Lee-Read  (HLR) composite Fermi liquid (CFL) theory \cite{HLR} furnished a good venue to study this $p$-wave pairing state \cite{lyw} but the non-relativistic nature of HLR theory and the breakdown of the particle-hole symmetry (PHS)\cite{kivelson,ldh} block to clarify this ambiguity.  Therefore, a relativistic CFL theory is eagerly called together with the following facts: The PHS of Jain's sequences of FQHE \cite{jain};  the PHS wave function for $\nu=\frac{5}2$ state shown by numerical calculations \cite{RH};  as well as the requirement of the universal origin for $\frac{1}2$ anomalous Hall conductivity in $\nu=\frac{1}2$ CFL \cite{kivelson,ldh}.  

Recently, the  enthusiasm of research for reexamining the CFL theory is aroused by experiments \cite{exp,exp1,exp2}.  A careful experiment of the composite fermion (CF) Fermi wave vector measurement through commensurability effects in the presence of a periodic grating suggests the breakdown of the PHS in FQHE \cite{exp} while theoretical explanation for the experiment must arise from PHS models \cite{jain1,fisher1}. Furthermore, the PHS breaking CFL state is not energetically favored as shown by numerical simulations \cite{ger}.

Two proposals were newly made in order to try to reveal this PHS enigma.  Barkeshli et al construct an anti-CF theory which is particle-hole conjugate to HLR theory \cite{fisher1}. Son's dual neutral Dirac CF (DCF) theory is based on a duality between a charged free Dirac fermion with a single cone in an external magnetic field and a neutral DCF coupled to a gauge field, a 2+1 electrodynamics (QED$_3$) \cite{dtson}. A semion-anti-semion bound state interpretation of CF for Son's model was put forward \cite{cwang1,cwang2}.  Several subsequent works appear \cite{met,met1,ger,murthy,mross,ors,ors1,pot,wanyang}.  An application of this duality to the surface state of a 3+1 dimensional topological insulator was  described \cite{met,met1} and the analogy to the CF of half-filled Landau level was exploited \cite{ger}. A Hamiltonian version of CF theory was updated to a PHS one \cite{murthy}.  An explicit derivation of Son's duality was provided \cite{mross}.  For more recent progresses, see a latest review \cite{sonre}.

The CF in FQHE is a composite object of an electron with attached even number of flux quanta \cite{jain}. Based on the observation that the external magnetic field is exactly cancelled at the half-filled Landau level by the average value of a fictitious statistical magnetic field which arises from the flux attachment, HLR \cite{HLR} developed the CFL theory near $\nu=\frac{1}2$.  The success of explaining many experiments \cite{Will,kane,gold} indicated the powerfulness of HLR theory. The energy gap from activation energy measurement  which is linearly proportional to the reduced residual magnetic field shows the existence of the CF Landau levels \cite{du}.  With a CFL, the Moore-Read Pfaffian state  \cite{MR} becomes natural because it is nothing but a $p_x+ip_y$-wave pairing states of CFs \cite{GR}. However, the lacking of the PHS leads to the anti-Pfaffian, the particle-hole conjugation of Pfaffian, is not  simply defined in HLR's framework \cite{lev,lee,fisher1}.    

There seems a barrier between the microscopic model and the CFL theory: The free flux attached CF has the same mass as the band mass of electron while the CFL theory phenomenologically replaces it with an effective mass which is in Coulomb energy scale \cite{HLR}. The  effective mass cannot be obtained by the declared mass renormalization in a perturbative calculation of the CFL theory. Later a Hamiltonian formalism calculation \cite{murthy2} and the temporal gauge calculation in one-loop level \cite{yu} can have a cancellation of the band mass while an artificial cut-off was introduced. These hint a need for a nonperturbative method. 

Learning from the recent progresses and the seminal HLR theory, we see that a DCF theory has  a priority for a PHS theory.  But
the questions waiting for replying are: (i) Can we give a straightforward relation between the  DCF model and a two-dimensional non-relativistic interacting electron gas in a strong external magnetic field?  (ii) How can we have a DCF whose Landau level gap $\propto B^*$ as measured  experimentally \cite{du}, instead of the well-known $\propto \sqrt {B^*}$ ? 

In this paper, we try to give such a DCF model at the mean field level that may answer these two questions from the interacting CFL theory.  Our starting point is the mean field theory of the composite fermion field theory.  The  spinless non-relativistic CFs are subject to a residual magnetic field and can be transformed into a Dirac fermion but the pseudospin-down component plays a role of an auxiliary particle with no dynamics \cite{ng}. 

In fact, if the CF wave function is in the $\nu^*$-th CF Landau level, the auxiliary particle wave function is in the  $\nu^*-1$-th CF Landau level, which implies the mixing of adjacent FQH states (or the mixing between adjacent DCF Landau levels).  The  interaction between CFs induces the dynamics of the auxiliary particle. As a result, a modified DCF theory emerges. We show that this emergent DCF theory is perturbatively unstable in the sense that the DCF collapses in a weak repulsive interaction while it is stable when the interaction is strongly repulsive. With the help of existed numerical data, we find that the DCF model applied to Jain's sequence is most likely stable and the model parameters such as the "speed of light" of the theory can be fixed.  We then build a DCF theory which reveals the enigma of PHS.  

This paper is organized as follows: In Sec. \ref{II}, we will propose a modified Dirac equation to show the nonperturbative emergence of the  DCF.  In Sec. \ref{III}, we will discuss the consequences of this DCF theory. We will show that the DCF Landau level gap is proportional to the effective magnetic field $B^*$, the duality between DCF and QED$_3$ and the stability of Jain's sequence are also studied. Sec. IV is our conclusions.

\section{Emergence of Dirac composite Fermions}\label{II}

 The CFL theory we would like to reformulate is described by the Hamiltonian 
\begin{eqnarray} 
H_{CF}&=&\frac{1}{2m_b}\sum_i[-i\hbar\nabla_i+\frac{e}{c}{\bf A}({\bf x}_i)-\frac{e}{c}{\bf a}({\bf x}_i)]^2\nonumber\\
&+&\sum_{i<j}V({\bf x}_i-{\bf x}_j),
\end{eqnarray}
where $V({\bf x})=\frac{e^2}{\varepsilon |{\bf x}|}$ is the Coulomb interaction and the neutralized background 
potential is omitted. The statistical gauge field ${\bf a}({\bf x}_i)$ is the gradient of the singular phase of the many body CF wave function and $\nabla\times {\bf a}=\frac{4\pi}{e/\hbar c }\sum_i\delta({\bf x}-{\bf x}_i)$.
For GaAs, the dielectric constant $\varepsilon=12.6$ and the electron band mass $m_b=0.07 m_e$. 
  
We would like to study the FQHE in the lowest Landau level of electrons. Our starting point is the mean field approximation which means that $ {\bf a}$ is approximated by $\bar {\bf a}$ that satisfies
 $
 \nabla\times{\bf \bar a}=B_{\frac{1}2},
$ 
 where $B_{\frac{1}2}$ is a magnetic field corresponding to a half-filled Landau level, i.e., $ 2\pi l^2_{B_{\frac{1}2}}\rho_e=\frac{1}2$; $l_B=\sqrt{\hbar c/eB}$ is the magnetic length. 
The mean field Hamiltonian reads
 \begin{eqnarray}
H_{MF}&=&
 \frac{1}{2m_b}\sum_i{\bf P}_i^2+\sum_{j< i}V({\bf x}_i-{\bf x}_j)\nonumber\\
 &=&\sum_i h_0({\bf x}_i)+\sum_{j<i}V({\bf x}_j-{\bf x}_i),  \end{eqnarray} 
where ${\bf P}=-i\hbar\nabla_i+(e/c){\bf A}^*$ with ${\bf A}^*={\bf A}-\bar{\bf a}$; $\nabla\times {\bf A}^*=B^*$ is the reduced residual magnetic field. 

 We look at the single CF problem:
   \begin{eqnarray}
h_0\varphi({\bf x})=E_{NR}\varphi({\bf x}).\label{NR}
 \end{eqnarray}
A special solution is $\varphi=\chi$, a CF wave function of the $\nu^*$ Landau level. Jain's sequences \cite{jain} give rise to the electron Hall coefficients found in experiments  
\begin{eqnarray}
\nu=\nu_{particle}=\frac{\nu^*}{2\nu^*+1},~\nu=\nu_{hole}=\frac{\nu^*+1}{2\nu^*+1}, \label{jain}
\end{eqnarray}
with $\nu^*=1,2,\cdots$. The former is $\nu<\frac{1}2$ sequence for electrons and the latter is $\nu>\frac{1}2$ for holes. The meaning of Jain's sequence is that the integer quantum Hall effects of CFs correspond to the FQHE of electrons. 

\subsection{ DCF equation} 

{As mentioned in the introduction, the DCF theory has a natural advantage in solving the problems encountered in the HLR theory, such as the particle-hole symmetry and the $\frac{1}{2}$ anomalous Hall conductivity in $\nu=\frac{1}{2}$ CFL\cite{kivelson,ldh}. Therefore we would like to propose a DCF theory with the property that it reduces to the non-relativistic CFL in some limit of the parameters.} Using (2+1)-dimensional gamma matrices $\gamma^0=\sigma^z$, $\gamma^1=\sigma^z\sigma^x=i\sigma^y$ and $\gamma^2=\sigma^z\sigma^y=-i\sigma^x$ where $\sigma^a$ are Pauli matrices,  we examine the following modified Dirac equation,
\begin{eqnarray}
[\mathbb{C}\gamma^0i\hbar\frac{\partial}{\partial t}-v\gamma^aP_a-2m_bv^2]\psi(x,t)=0,\label{CFD}\end{eqnarray}
where  $\mathbb{C}={\rm diag}(1-C,-C)$ is a 2$\times 2$ diagonal constant matrix and the pseduo-spinor $\psi({\bf x},t)=e^{-iEt/\hbar}\psi({\bf x})=\left(\begin{array} {c}\chi\\\phi\end{array}\right)$;  $v$, different from that in Ref. [\onlinecite{ng}], may not be the genuine speed of light $c$ but a constant with a dimension of speed (see below).  When $C=0$,  $\phi$ is an auxiliary field with no dynamics \cite{ng}, i.e.,
\begin{eqnarray}
\phi=P_+\chi/(2m_bv) \label{phi} \label{lower},
\end{eqnarray}
where $P_\pm=P_x\pm iP_y$ and $\chi$ obeys
$$E\chi=\frac{1}{2m_b}(p_x^2+p_y^2-\frac{\hbar eB^*}{c})\chi+2m_bv^2\chi.$$
This is the non-relativistic Schrodinger equation (\ref{NR}) but $E_{NR}=E-2m_bv^2$. As $P_+$ is the lowering operator of the  CF Landau level,
if $\chi$ is the CF wave function of the $\nu^*$-th CF Landau level,  the auxiliary field $\phi$ is the CF wave function in $\nu^*-1$-th CF Landau level (see eq. (\ref{lower})). Therefore, if we only count the interaction {within the} intra-Landau level and adjacent Landau levels of CFs, the interaction can be approximated as
\begin{eqnarray}
\sum_{j<i}V({\bf x_i}-{\bf x}_j)\approx\sum_{{\bf x},{\bf x}'} \psi^\dag({\bf x})  \psi^\dag({\bf x}') \frac{e^2}{\varepsilon |{\bf x}-{\bf x}'|} \psi({\bf x}')\psi({\bf x}).\label{int}
\end{eqnarray}
 
\subsection{ Dynamical $\phi$ field} Notice that because Jain's sequence (\ref{jain}) does not include $\nu=1$ integer quantum Hall effect, $\nu^*=0$ will not be taken by $\chi$, i.e., $\phi$  may not meaningless and at least is the wave function of the CF lowest Landau level. (Do not confuse with the lowest Landau level of electrons). The interactions between $\phi$ as well as between $\phi$ and $\chi$ are also in the order of Coulomb potential.  In the lowest Landau level of electrons,  all electron's dynamics can come from the interaction.  Namely, the interaction (\ref{int}) can supply $\phi$ {with} dynamics, i.e., $C\ne 0$ in Eq. (\ref{CFD}). The  scale of a CF energy $E$ is of the order of the Coulomb scale $\frac{e^2}{\varepsilon l_B}$, the average  Coulomb potential per particle.  We thus take a mean field approximation for the interaction 
 \begin{eqnarray}
 \sum_{{\bf x}'} \psi^\dag({\bf x})  \psi^\dag({\bf x}') \frac{e^2}{\varepsilon |{\bf x}-{\bf x}'|} \psi({\bf x}')\psi({\bf x})\approx \psi^\dag({\bf x} )CE\psi({\bf x} ),\label{8}
 \end{eqnarray}
 i.e., we use $CE$ to approximate the interaction potential that a CF feels and assume it differs a real number factor $C(\nu)$ from $E$.  In this way, $\phi$ becomes dynamical. As $P_\mp$ raises and lowers DCF Landau level, the "speed of light" $v$ couples the wave functions in adjacent DCF Landau levels and then it reflects the strength of the adjacent Landau level mixing of DCFs.  Also because of in the lowest Landau level of electrons,  $v$ is governed by the Coulomb scale, i.e., one can take
 \begin{eqnarray}
 v\equiv\frac{D(\nu)\hbar}{m_cl_B}=D(\nu)\frac{\alpha}\varepsilon c,\label{9}
 \end{eqnarray}
  with $m_c=\varepsilon \hbar^2/e^2l_B$ being the Coulomb mass and $\alpha=\frac{e^2}{\hbar c}\approx \frac{1}{137}$ the fine structure constant. The filling factor dependent constants $C(\nu)$ and $D(\nu)$ will be determined later.

 \subsection{ Nonperturbative  DCF }
 
  We now are ready to solve Eq. (\ref{CFD}). Writing the equation as
 \begin{eqnarray} 
&&(1-C)E\chi=\frac{v \hbar}{l_{B^*}}\sqrt 2a^\dag\phi+2m_bv^2\chi,\label{10}\\
&&-C E\phi=\frac{v\hbar}{l_{B^*}}\sqrt 2a\chi-2m_bv^2\phi\label{11},
 \end{eqnarray}
 where we have taken the symmetric gauge ${\bf A}^*=(\frac{B^*}2y,-\frac{B^*}2x)$ with $B^*$ in the negative $z$-direction; $l_{B^*}=\sqrt{\hbar c/eB^*}$ and $z=(x+iy)/l_{B^*}$. The lowering and raising operators of the CF Landau levels are given by {
 $
 a=-i(\partial_{\bar z}+\frac{z}2)/\sqrt 2,a^\dag=-i(\partial_{z}-\frac{\bar z}2)/\sqrt 2,
$
 with $[a,a^\dag]=1$. Though the interaction induced $\mathbb{C}$ matrix in Eq. (\ref{CFD}) may damage the hermiticity of its Hamiltonian, we will show that when the energy is real, the corresponding eigenstates are orthogonal, norm unity, complete and closed, namely, they share the same properties as the eigenstates of an hermitian Hamiltonian. Besides this nice property, other non-hermitian systems have been studied theoretically and experimentally\cite{dk}. Therefore we believe the non-hermitian Hamiltonian under consideration is not an obstacle as long as we remain in the real energy region. } 
 
{
Substituting Eq. (\ref{11}) into Eq. (\ref{10}), we obtain an algebraic equation for the spin component $\chi$,
\begin{equation}
\frac{l^{*2}_b}{(v\hbar)^2}(2m_bv^2-CE)[(1-C)E-2m_bv^2]\chi=2a^\dagger a\chi.
\end{equation}
Therefore the eigen wave function of $\chi(z;\nu^*)$ is the same as that of the $\nu^*$-th Landau level, namely, 
\begin{equation}
\chi(z,\nu^*)=\frac{1}{\sqrt{\nu^*!}}(a^\dagger)^{\nu^*}\chi(z,0),
\end{equation}
with
\begin{equation}
\chi(z,0)=\frac{1}{\sqrt{2\pi}l_B^*}e^{-\frac{z^*z}{4}}.
\end{equation}
From Eq. (\ref{11}), we know that $\phi\propto a \chi$, and then  the solutions of Eq. (\ref{CFD}) are of the form,
\begin{equation}
\left(\begin{array} {c}\chi\\\phi\end{array}\right)=\frac{1}{\tilde{N}}\left(\begin{array} {c}\chi(z,\nu^*)\\b\chi(z,\nu^*-1)\end{array}\right),\label{wavefunction}
\end{equation}
with $\tilde{N}$ being the normalizing factor and $b = \frac{\sqrt{2\nu^*}v\hbar}{(2m_bv^2-CE(\nu^*))l_b^*}$ is some constant given by Eq. (\ref{11}). In terms of the structure of the eigen wave functions, the orthogonality, completeness, and closeness are obvious.}

 Defining $\omega^*_c=\sqrt{eB^*c/\hbar}$ which is the relativistic cyclotron motion frequency and $\tilde\omega^*_c=\frac{v}c\omega^*_c$, the CF Landau levels are determined by
\begin{eqnarray}
E(\nu^*)&=&\pm\sqrt{\frac{2\nu^*\hbar^2\tilde\omega_c^{*2}+M_b^2v^4}{C(C-1)}}
-\frac{m_bv^2}{C(C-1)}, \label{DCFLL}
\end{eqnarray}
where{ $a^\dagger a \chi =\nu^*\chi$ }and $M_b^2=\frac{(2C-1)^2}{C(C-1)}m_b^2$.
We see that $E$ diverges at $C=0,1$ and is real only when $$C<\frac{1}2-\frac{1}2\sqrt{1-\frac{4m_b^2v^2}{4m_b^2v^2+2\nu^*\hbar^2\tilde\omega_c^{*2}}}$$ or $$C>\frac{1}2+\frac{1}2\sqrt{1-\frac{4m_b^2v^2}{4m_b^2v^2+2\nu^*\hbar^2\tilde\omega_c^{*2}}}.$$ In the zero band mass limit ($m_b\to 0$), this implies that the DCF $\psi$ is not stable for $0\leq C\leq 1$.  With a weak 
repulsive interaction, the dynamic DCF collapses to a non-relativistic CF. In this sense, the DCF can only emerge nonperturbatively.

\section{Consequences of DCF}\label{III}
\subsection{ Gap  $\propto B^*$} In the lowest Landau level of electrons, the electron cyclotron motion energy is much larger than the Coulomb energy scale. Notice that both factors, $v$ and $\omega^*_c$, in $\tilde \omega^*_c$ are proportional to $\sqrt{B^*}$. In the zero band mass limit, hence, the CF Landau level energy (\ref{DCFLL}) tends to  
\begin{eqnarray}
E=\pm \hbar\nu^*\frac{eB^*}{F m_cc}.~\label{zml}
\end{eqnarray}
where $F(\nu^*)=\frac{\sqrt{\nu C(C-1)}}{\sqrt{2}D}$ so that $m^*=F(\nu^*)m_c$ is the CF effective mass. \cite{HLR,du,morf}. Thus, the DCF Landau level gap is basically linearly dependent on $B^*$. This shows a crucial difference between the spectrum of this DCF and a conventional Dirac fermion Landau level which is proportional to the square root of the external magnetic field, $\sqrt{B^*}$. Experimentally,  the energy gaps from activation energy measurements is $\Delta \sim \hbar\frac{eB^*}{m^*c}-\Gamma$ with a small broadening factor $\Gamma\sim 2$K \cite{du}.

\subsection{Dual to QED$_3$} We rescale $\chi$, $\phi$, and $v$ by
\begin{eqnarray}
\phi'=G^{-1}\phi, \chi'=G\chi ,
 v'=-\frac{v}{\sqrt{C(C-1)}}\nonumber,  
 \end{eqnarray}
for a real  $G=[(C-1)/C]^{1/4}$  
 and eq.(\ref{CFD}) becomes
 \begin{eqnarray}
E\chi'&=&v'P_-\phi'+2Cm_bv'{^2}\chi', \nonumber\\
E\phi'&=&v'P_+\chi'-2(C-1)m_b v'{^2}\phi'.  \label{prime}
\end{eqnarray} 
 If we take $v'^\mu=(1,v',v')$, eq. (\ref{prime}) is nothing but the free Dirac equation with an external field,
\begin{eqnarray}
[\gamma^\mu v'{^\mu}(i\hbar \partial_\mu-\frac{e}cA{^*}_\mu)-\mathbb{M}_bv'{^2}]\psi'=0. \label{dual}
\end{eqnarray}
where the mass matrix $\mathbb{M}_b=-2m_b{\rm diag}(C,C-1)$. In the zero band mass limit, as the
amount of recent researches showed,  the free Dirac equation (\ref{dual}) is dual to a QED$_3$ for a neutral DCF $\tilde\psi~$ \cite{dtson,cwang1,met,met1,ger,murthy,mross}, whose Lagrangian is given by
\begin{eqnarray}
\tilde{\cal L}&=&\int d{\bf x}dt[\bar{\tilde\psi}\gamma{^\mu}c^\mu(i\hbar\partial_\mu
+\frac{g}c\tilde a_\mu)\tilde\psi\nonumber\\
&+&\frac{eg}{4\pi \hbar c}\epsilon_{\mu\nu\rho}A^{*\mu}\partial ^\nu\tilde a^\rho+\cdots].\nonumber
\end{eqnarray}
where $c^\mu=(1,c,c)$ and $g$ is coupling constant of the QED$_3$; $"\cdots"$ includes the Maxwell term of $\tilde a$ with a $v'$-dependent coupling constant \cite{mross}.

\subsection{ Stability for Jain's sequence}  Our theory is a mean field approximation of the microscopic model of the two-dimensional electron gas in a strong magnetic field. There are two model parameters, $C$ and $D$.  While the former connects the mean field CF Coulomb energy with the mean field energy per DCF, the latter is essentially equivalent to the CF effective mass $m^*$. To self-consistently determine the parameters, one needs a couple of mean field equations. We can use the DCF wave function to calculate the DCF Coulomb potential and then let it relate to the DCF energy (\ref{DCFLL}). This obtains one of mean field equations. Solving this equation gives rise to a relation between $C$ and $m^*$.  However, due to the strongly correlated nature, it is difficult to get another mean field equation to solve this relation. On the other hand,  there were many numerical calculation results of the ground state energy and the effective mass of the CF for the microscopic model \cite{morf,morf1,wan,RR, YCH, HS,Ao,ZSH,Morf2}. Thus, we can use these presented numerical calculation results to input either $C$ or $m^*$ and then the other one is determined by the mean field equation.

Applying the mean field approximation to Jain's sequences, we find that with the help of these numerical data, Jain's sequences are most likely stable, at least for $\nu=\frac{1}3,\frac{2}5,\frac{3}7$ and $\frac{4}9$.  This is what we will do in this subsection.

{Using our mean field approximation (\ref{8}), the energy $E$ and the typical Coulomb potential per DCF is related to one another by
\begin{eqnarray}
|E|\sim \langle \frac{e^2}{\epsilon |\bf x|}\rangle/C, \label{EC}
\end{eqnarray}
where $\langle \frac{e^2}{\epsilon |\bf x|}\rangle$ is the expectation value of the Coulomb potential for the CF eigen states. In the zero band mass limit $m_b\to 0$, the eigen wave function (\ref{wavefunction}) becomes,
\begin{equation}
\left(\begin{array} {c}\chi\\\phi\end{array}\right)_{\nu^*}=\sqrt{\frac{C-1}{2C-1}}\left(\begin{array} {c} -\sqrt{\frac{C}{C-1}}\chi(z,\nu^*)\\\chi(z,\nu^*-1)\end{array}\right).
\end{equation}
Therefore the expectation value of the mean field Coulomb potential of the single particle wave function can be approximated as,
\begin{eqnarray}
 \langle \frac{e^2}{\epsilon |\bf x|}\rangle_{\nu^*}&=&(\frac{C}{2C-1}\frac{(2n-1)!!}{2^nn!}+\nonumber\\
 &&\frac{C-1}{2C-1}\frac{(2n-3)!!}{2^{n-1}(n-1)!})\sqrt{\frac{\pi}{2}}\frac{e^2}{\epsilon l_B^*} \label{CC}
\end{eqnarray}
Eqs, (\ref{EC}) and (\ref{CC}) give rise to a relation between $E$ and $C$ for a given $\nu^*$, which are plotted in Fig. \ref{dcf-eg} for the first three $\nu^*$ and $C>1$. We also notice that if we determine $C$ according to $E$, $C$ becomes complex if $E$ exceeds the maximal magnitude in Fig. \ref{dcf-eg} for a given $\nu^*$, say, $E>0.36\frac{e^2}{\epsilon l_B}$ for $\nu^*=1$.
\begin{figure}
 	\includegraphics[width=0.45\textwidth]{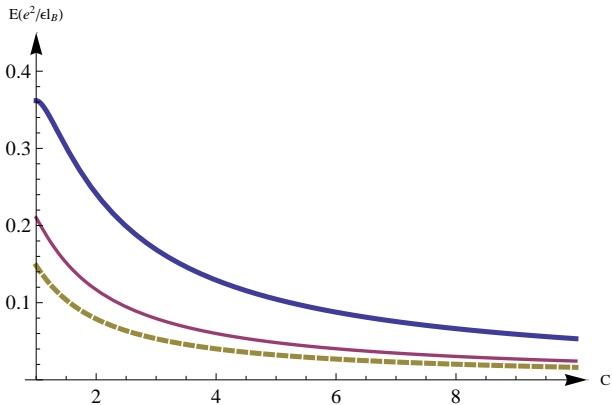}
 	\caption{(Color online)
 	The energy $E(C)$ is determined through Eqs. (\ref{EC}) and (\ref{CC}) for the real $C$. The blue(thick) line is for $\nu^*=1$, the purple(solid) line is for $\nu^*=2$, and the yellow(dashed) line is for $\nu^*=3$. 
 	}
 	\label{dcf-eg}	
 \end{figure}
We can also estimate $E(C)$ through Eq. (\ref{zml}) if the CF effective mass is inputted.  We use the numerical estimation to the CF effective mass for the $\nu^*$th Landau level by Morf et al \cite{morf}
\begin{equation}
m^*(\nu^*)=\frac{\hbar^2\epsilon}{e^2 l_B}\frac{2}{\pi}(\ln(2\nu^*+1)+4.11).
\end{equation}
 The intersection between these two $E(C)$ curves determines $C(\nu^*)$ for a given $\nu^*$.  
Taking $\nu^*=1$ as an example, the effective mass data leads to the intersection is at $E=0.101\frac{e^2}{\epsilon l_B}$ and $C=5.21$. For the filling factor $\nu=\frac{2}5,\frac{3}7$ and $\frac4 9$, the corresponding values of $E(\nu^*),C(\nu^*)$ are listed in Table I (marked by $\&$, the dielectric constant $\varepsilon=12.6$ for GaAs). The  corresponding magnitudes of  $E_g$, $D$ and the Fermi velocity $v$ are also calculated. We see that all values of $C$ for these filling factors are real and larger than 1. This indicates the stability of the Jain's sequences in this DCF mean field theory. For $C<1$, Eq. (\ref{zml}) gives an imaginary $E$ which does not coincide with Eq. (\ref{EC}) and means that there is no such a mean field solution. }

On the other hand, if we know the ground state energy $E_g$ and take $E=E_g$ in Eq. (\ref{EC}), $C$ can be determined by solving
 \begin{equation}
 \langle \frac{e^2}{\epsilon\bf x}\rangle_{\nu^*}/|C|\sim |E_g|.\label{23}
 \end{equation}
{For example, if we use the HRL non-relativistic energy gap $\Delta(\nu)$ to estimate $E_g$\cite{HLR}, say  for $\nu=\frac{1}{3}$ (or $\nu^*=1$),
 \begin{equation}
 E_{g}(\nu^*=1)=\Delta(\nu=\frac{1}{3})=0.1\frac{e^2}{\varepsilon l_B},
 \end{equation}
 where $\Delta(\nu=\frac{1}{3})$ is a numerical result chosen from Ref. \cite{morf}. Therefore the corresponding $C(\nu=\frac{1}{3})=5.24$. 
 Many numerical calculations for the ground states existed \cite{morf,HS,Ao,ZSH,Morf2}.}
 We list our calculation results of the parameters $C,F,$ and $D$ as well as the Fermi velocity according to the existing numerical data of $E_g$ in Table I.  Notice that some of $C$ are complex number because $E_g$ is too large as explained before. This indicates that either the estimation 
 of $C$ through   Eq. (\ref{23}) merely is not a reliable way, or the mean field theory is not stable.

 \begin{center}
\begin{tabular}{||c|c|c|c|c|c||}
\hline$~\nu~$&$E_g /(\frac{e^2}{\varepsilon l_B})$&$~~C~~$&$~~F~~$&$~~D~~$&$~~|v|/c~~$\\ \hline
$\frac{1}3$&$0.10^*$& 5.23& 3.33& 0.58 &$  3.34\times 10^{-4}$\\ \hline
$\frac{1}3(\&)$&$0.101 $& 5.21&3.32$^*$& 0.58 &$ 3.34\times 10^{-4} $\\ \hline
$\frac{1}3$&$0.41^{\diamond}$&complex &$0.81 $ &X&X \\ \hline
$\frac{1}3$&$0.41^{\clubsuit}$&complex&$0.81 $ &X &X \\ \hline
$\frac{1}3$&$0.41^{\heartsuit}$&complex&$0.81 $ &X&X \\ \hline
$\frac{1}3(\&)$&$0.05 $ &$10.7 $ &$ 6.67^{\spadesuit}$ &$0.62 $ &$  3.61\times 10^{-4}$ \\ \hline
$\frac{2}5$&$0.15^*$&$1.52 $&$2.67 $&$0.15 $&$ 0.86\times10^{-4} $\\ \hline
$\frac{2}5(\&)$&$0.11 $&$2.13 $&$ 3.64^*$&$0.19$&$1.10\times 10^{-4} $\\ \hline
$\frac{2}5$&$0.43^{\heartsuit}$&complex&$0.93 $ &X &X \\ \hline
$\frac{2}5$&$0.06 $ &$ 3.74$ &$ 6.25^{\spadesuit}$ &$0.23 $ &$ 1.33\times 10^{-4}$ \\ \hline
$\frac{3}7$&$0.185^*$&complex& 2.32 & X&X\\ \hline
$\frac{3}7(\&)$&$0.11 $&1.39 & 3.86$^*$& 0.09 &$ 0.51\times10^{-4} $\\ \hline
$\frac{3}7$&$0.44^{\heartsuit}$&complex&$0.97 $ &X & X \\ \hline
$\frac{3}7(\&)$& $0.07 $&$2.17 $ &$5.88^{\spadesuit}$ &$0.13 $ &$ 0.73\times 10^{-4}$ \\ \hline
$\frac{4}9$&$0.212^*$&complex& 2.10 & X&X\\ \hline
$\frac{4}9(\&)$&$0.11 $&$1.04 $& 4.02$^*$& 0.024 &$ 0.14\times10^{-4} $\\ \hline
\end{tabular}
\end{center}

Table I: {* are taken from [\onlinecite{morf}],
 $\diamond$ are from Fig. 2 and Fig. 4 of [\onlinecite{HS}], $\spadesuit$ are from Fig. 2 of [\onlinecite{Ao}], $\clubsuit$ are from Fig. 2 and Fig. 3 of [\onlinecite{ZSH}], and $\heartsuit$ are from Table I of [\onlinecite{Morf2}]. $|v|/c$ is determined through Eq. (\ref{9}). "complex" means that $C$ has a nonzero imaginary part, i.e., for the corresponding numerical data, there is no stable solution in this $C$. The notion "\&" in the first column stands for that the $C$ in that row is determined from the effective mass, otherwise it's from the ground state energy.} 
 
 We summarize and discuss the previous results: 
 
 (i) Although $C$  in Table I are only rough estimation for Jain's sequence $\frac{\nu^*}{2\nu^*+1}$ with $\nu^*=1,2,3,4$, we see that many numerical results support that the DCF is stable in the mean field approximation because the magnitudes of $C$ are real and larger than 1. Since these states are gapped, the Chern-Simons gauge fluctuation and the residual interaction will not severely alter these mean field results. 
 
 (ii) For a given $\nu^*$,  while the effective mass to estimate $C$ gives a real number for $\nu^*=1,...,4$, it may be complex by using the ground state energy. In fact, both methods to determine $C$ may false if the numerical magnitude of the energy is too large so that it exceeds the maximum given by Fig. \ref{dcf-eg}. With the former, the requirement is to match two mean field energies (\ref{EC}) and (\ref{zml}) while the mean field energy (\ref{EC}) is directly identical to the numerical ground state energy with the latter. Obviously, the former way should be more consistent.   Moreover,
in the numerical calculations of the ground state energy, there were many uncertainty conditions to confine the precision of the numerical data such as the type of the interactions, the finiteness scalings, the boundary conditions and so on .     
   
  (iii)  Due to the PHS of the
 Dirac equation, the
 particle-hole transformation gives rise to
\begin{eqnarray}
&&(1-C)E\chi_h=vP_+\phi_h+2m_bv^2\chi_h, \nonumber\\
&&-CE\phi_h=vP_-\chi_h-2m_bv^2\phi_h. 
\end{eqnarray} 
The difference from the particle's equation is only  in exchanging $P_+\leftrightarrow P_-$. This results in $E(\nu^*)\to E(\nu^*+1)$ in eq. (\ref{DCFLL}) and gives the Jain's sequence $\frac{\nu^*+1}{2\nu^*+1}$ for hole. We expect our theory is also stable because the experimental data showed a nearly symmetric CF Landau level gap between $\nu$ and $1-\nu$ \cite{du}. 

{ (iv) In the estimation of the Coulomb energy $\langle \frac{e^2}{\epsilon |\bf x|}\rangle$, we only considered single particle contribution. If we take the many body effects into count, then it will change the expectation values of the Coulomb energy, thus change $C$. For example, if we consider a two-body wave function,
	\begin{equation}
	\Psi_1=\psi_n(x_1)\psi_m(x_2),\quad \Psi_2=\psi_m(x_1)\psi_n(x_2),
	\end{equation}
where $\psi_n$ is the wave function for the nth Landau level. The Coulomb energy between $x_1$ and $x_2$ is $e^2\int \frac{1}{r_{12}}|\psi_n|^2|\psi_m|^2dr_1dr_2$. In our mean field approximation, we only considered the case when $n=m$. If $n\neq m$, this term will increase the Coulomb energy and  then enlarges $C$. The exchange energy is $e^2\int \frac{1} {r_12}\psi_n^*(x_1)\psi_m(x_1)\psi_n(x_2)\psi_m^*(x_2)d r_1dr_2$, which in general will decrease the Coulomb energy and gives  a smaller $C$. Besides the many body effects, we also assumed zero band mass limit $m_b\rightarrow 0$. If the band mass is small but non-zero, it will also increase $C$. These uncertainties will leave to further studies.}

 (v) For $\nu^*=\infty$, i.e., $\nu=1/2$, the system becomes gapless and $C\to 1/2$ by using the effective mass $m^*(\nu=1/2)$.  Then, the mean field theory is not stable for $\nu=1/2$ and $5/2$. However, we emphasize that the mean field estimation of $C$ in these filling factors may be altered by the strong gauge fluctuation.  We thus expect the DCF theory might still work for these even denominator filling factors. Further study is required.

\subsection{ Anomalous  Hall conductivity $-\frac{1}2\frac{e^2}h$} 
For a given $\nu$, we have fixed $C$ and $v$ with the ground state energy or the effective CF mass. However, the topological properties will not change as $C$ varies if we keep $C>1$ . For a large $C$, $C-1\approx C$, we obtain the standard Dirac equation,
$$
(i\tilde v^\mu\gamma^\mu D_\mu-\tilde m_b \tilde v^2)\psi=0,
$$
where $\tilde v^0=1, \tilde v^a=\tilde v= -v/C$ and $\tilde m_b=-m_b/C$. $\tilde m_b$ has an opposite sign to $m_b$. Although we do not yet prove this DCF is stable for $\nu=1/2$, there is an axial anomaly and then an anomalous Hall effect $\sigma^{CF}_{xy}=\frac{{\rm sgn}(\tilde m_b)}2\frac{e^2}h=-\frac{1}2\frac{e^2}h$ once $C>1$ is real for $\nu=1/2$, . One can also arrive at this consequence according to Eq. (\ref{dual}).

 \section{Conclusions } We presented a nonperturbatively emergent DCF theory from the CFL theory. In this strong correlated theory, the PHS of FQHE, i.e., $\nu$ and $1-\nu$ symmetry of Jain's sequences, is restored. We showed that this DCF is most likely stable for Jain's sequence.  The energy gap is linearly dependent on the effective residual magnetic field for the CFs.  The dual to Son's QED$_3$ was proved and the mystery of minus one-half anomalous Hall conductivity was revealed.  We expect this DCF theory is stable for $\nu=1/2$ or $\nu=5/2$ and then give rise to the origin of the anomalous  Hall conductivity for $\nu=1/2$ and the relativistic Majorana fermion in $\nu=\frac{5}2$. 

\acknowledgments

The authors thank Yong-Shi Wu for drawing their attention to  the latest progresses in this field.  This work was supported by  NNSF of China (11474061).

\end{document}